\documentclass[%
aps,
reprint,
superscriptaddress,
amsmath,amssymb,
prb,
floatfix,
]{revtex4-2}
\usepackage{hyperref}
\usepackage{graphicx}
\usepackage{physics}

\begin{document}

\title{Tuning Hydrogen Adsorption and Electronic Properties from Graphene to Fluorographone}

\author{Gabriel R. Schleder \href{https://orcid.org/0000-0003-3129-8682}{\includegraphics[scale=.05]{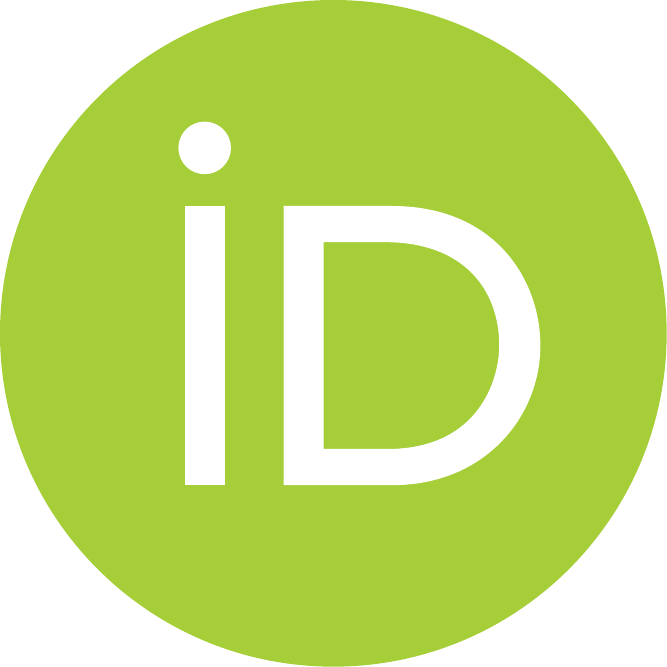}}}
\email{gabriel.schleder@ufabc.edu.br}
\affiliation{Federal University of ABC (UFABC), 09210-580 Santo Andr\'e, S\~ao Paulo, Brazil}

\author{Enesio Marinho Jr \href{https://orcid.org/0000-0003-4040-0618}{\includegraphics[scale=.05]{orcid_id}}}
\email{enesio.junior@ufabc.edu.br}
\affiliation{Federal University of ABC (UFABC), 09210-580 Santo Andr\'e, S\~ao Paulo, Brazil}

\author{Douglas J. R. Baquiao \href{https://orcid.org/0000-0002-4995-3799}{\includegraphics[scale=.05]{orcid_id}}}
\affiliation{Federal University of ABC (UFABC), 09210-580 Santo Andr\'e, S\~ao Paulo, Brazil}

\author{Yuri M. Celaschi \href{https://orcid.org/0000-0003-4209-3295}{\includegraphics[scale=.05]{orcid_id}}}
\affiliation{Federal University of ABC (UFABC), 09210-580 Santo Andr\'e, S\~ao Paulo, Brazil}

\author{Felipe Gollino \href{https://orcid.org/0000-0003-4921-7895}{\includegraphics[scale=.05]{orcid_id}}}
\affiliation{S\~ao Carlos Institute of Chemistry, University of S\~ao Paulo, 13566-590 S\~ao Carlos, S\~ao Paulo, Brazil}

\author{Gustavo M. Dalpian \href{https://orcid.org/0000-0001-5561-354X}{\includegraphics[scale=.05]{orcid_id}}}
\affiliation{Federal University of ABC (UFABC), 09210-580 Santo Andr\'e, S\~ao Paulo, Brazil}

\author{Pedro A. S. Autreto \href{https://orcid.org/0000-0002-3766-3778}{\includegraphics[scale=.05]{orcid_id}}}
\email{pedro.autreto@ufabc.edu.br}
\affiliation{Federal University of ABC (UFABC), 09210-580 Santo Andr\'e, S\~ao Paulo, Brazil}

\begin{abstract}
Graphene functionalization by hydrogen and fluorine has been proposed as a route to modulate its reactivity and electronic properties. 
However, until now, proposed systems present degradation and limited hydrogen adsorption capacity.
In this study, combining first-principles calculations based on density functional theory (DFT) and reactive molecular dynamics, we analyze the tuning of hydrogen adsorption and electronic properties in fluorinated and hydrogenated monolayer graphenes. 
Our results indicate that fluorine adsorption promotes stronger carbon--hydrogen bonds. 
By changing the concentration of fluorine and hydrogen, charge density transfer and electronic properties such as the band gap and spin-splitting can be tailored,
increasing their potential applicability for electronic and spintronic devices.
Despite fluorine not affecting the total H incorporation, the \textit{ab initio} molecular dynamics results suggest that 3\% fluorinated graphene increases hydrogen anchoring, 
indicating the hydrogenated and fluorinated graphenes potential for hydrogen storage and related applications.
\end{abstract}

\maketitle

\section{Introduction}
\label{S:1}
Graphene is a one-atom-thick layer of  \textit{sp}$^2$ bonded carbon atoms, which forms a two dimensional (2D) honeycomb lattice. This nanomaterial was first synthesized by Novoselov and  Geim \cite{Novoselov2004}, and since then has inspired the research and development of new 2D materials \cite{Jin2009,Lijie2010,Xu2013ChemRev,Liu2014,Zhu2015,Tao2015,Mannix2015,Feng2016,Alvarez-Quiceno2017a,Alvarez-Quiceno2017,Molle2017,Schleder-MLreview,PuthirathBalan2018,Schleder2019, padilha2019theoretical,Costa2019} for applications in nanoscale electronics.
Due to its unique and remarkable electronic \cite{CastroNeto2009}, thermal \cite{Balandin2011} and mechanical properties  \cite{Lee385}, graphene has been applied in several technological areas, such as flexible display screens \cite{Vlasov2017}, spintronic devices \cite{Han2014}, solar cells \cite{YiSong2016}, and electrochemical energy storage \cite{Raccichini2014}.

However, single-layer pristine graphene can be considered a zero-gap semiconductor (semimetal),  which hinders its application in technologies that require a semiconducting band gap such as optical and electronic devices. To overcome this issue, many experimental strategies have proposed the functionalization of graphene by reactions with organic and inorganic molecules, as well as chemical modifications of graphene surface \cite{Georgakilas2012}. Doping of graphene by ad-atoms adsorption, introducing point defects, applying external strains, or combinations of these, are some potential routes to open the band gap in graphene \cite{Widjaja2016,Kuila2012, vilasboas2019}. 

The adsorption of various chemical species on graphene is a widely studied subject \cite{Georgakilas2012,Pontes_2009}. Mohamed and co-workers investigated the application of nitrogen-doped graphene as a suitable photocatalyst \cite{Mohamed2018}. Poh \textit{et al.} proposed a method for the preparation of halogenated graphene by thermal exfoliation in gaseous halogen atmospheres \cite{Poh2013}. Chen and co-workers reviewed the properties of oxidized graphene and suggested the application of graphene-oxide in electrochemical studies and electroanalytical applications \cite{Chen2012}.

Another important functionalization strategy used is adsorbing hydrogen and/or fluorine atoms \cite{zhouPRB2010,Pumera2013,chronopoulos2017,feng2016review,Popov2013}.
Graphane is defined as a theoretical nonmagnetic semiconductor composed  by  100\%  hydrogenation  of  graphene, resulting in a CH stoichiometry \cite{Peng2014}. In addition, graphone is  the  midpoint between  graphene and graphane in which the graphene sheet is only one-side hydrogenated \cite{Zhou2009,Peng2014}. Experimental synthesis of graphone requires a back-side stabilization in order to overcome the thermodynamic and kinetic instabilities at room temperature \cite{vsljivanvcanin2012magnetism}. Usually this stabilization is carried out by metal surfaces such as Ir or Ni(111) which can also control the hydrogen atoms desorption (dehydrogenation) \cite{soni2018reactivity, zhao2015reversible}. 

\begin{figure*}[ht]
	\centering
	\includegraphics[width=\textwidth]{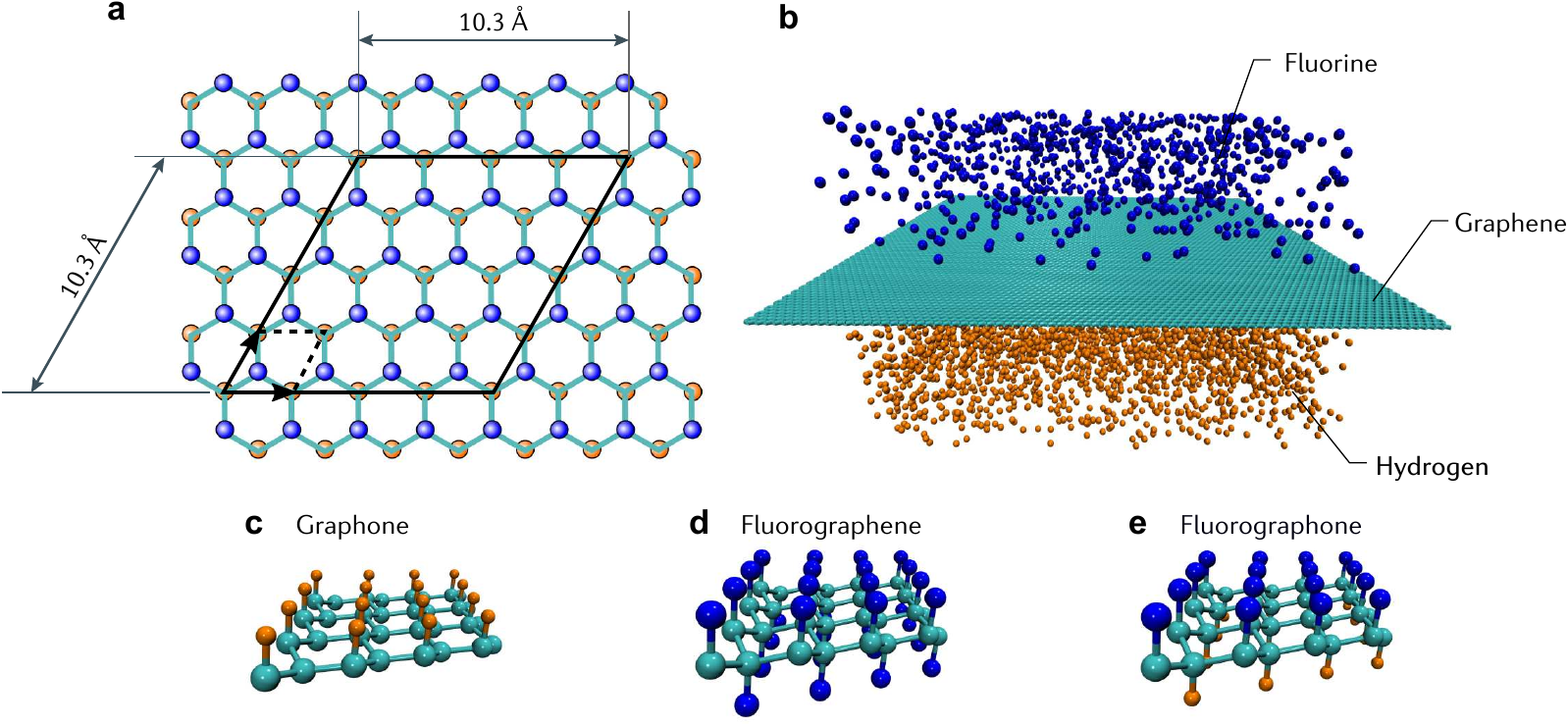}
	\caption{Configurations of graphone, fluorographene and fluorographone: (a) 4$\times$4 fluorographone supercell scheme (solid line) and unit cell (dashed line); (b) pristine graphene sheet with fluorine and hydrogen atmospheres used in the molecular dynamics simulations; 4$\times$4 supercells of (c) graphone, (d) fluorographene, and (e) fluorographone. Carbon, fluorine, and hydrogen atoms are represented in green, blue, and
orange, respectively}
	\label{fig:systems}
\end{figure*}

During the hydrogenation or fluorination processes, the planar \textit{sp}$^2$ structure of the pristine graphene changes to a tridimensional structure based also on \textit{sp}$^3$ hybridization \cite{PhysRevB-partiallyHFgraph}. These other configurations of the hydrogenated-graphene are classified according to the angles and the atomic positions, and are so-called chair or boat configurations \cite{Pumera2013}.
  
Recently, graphane was synthesized in a stoichiometric proportion similar to theoretical predictions, presenting intensive luminescence in the visible spectra and highly hydrophobic behavior \cite{Bousa2016}. Besides optoelectronic applications \cite{Jeon2011}, this can lead to chemically isolated structures where conducting graphene devices are surrounded by insulating fluorographene \cite{Lee2011}. 

Fluorographane,  which is the functionalized graphene by H and F incorporated in mixed atmospheres, was experimentally obtained \cite{Sofer2015} and showed heterogeneous electron transfer properties. 
The hydrophobic properties can be tuned under different atmosphere fluorine concentrations.
Fluorographane formation process and its properties are still subject to investigation. It is already known that fluorine atoms do not tend to form clusters \cite{Popov2013}, but in high concentrations, they can cause damage to the graphene sheet \cite{Paupitz2013}. The authors associate increased hydrogen incorporation to the presence of fluorine. 
Hydrofluorinated Janus graphene, a functionalized graphene with H and F adsorbed on opposite sides, has been proposed as a promising system in terms of piezoelectricity, reaching values one order of magnitude higher than single element adsorption \cite{Brazhe2013}, with potential adsorption properties \cite{jin2016self}.

To obtain physical insights regarding the challenge of reducing the material damage while further improving the hydrogen storage capacity, in the present work we investigate 
the hydrofluorinated Janus graphene, which here we describe as \textit {fluorographone}. 
Based on \textit{ab initio} calculations and reactive molecular dynamics, we systematically studied the modulation of hydrogen adsorption and the corresponding variation in physical properties of the hydrogenated and/or fluorinated graphene materials.
We compared the electronic and binding properties among different H and F functionalized systems
and our findings suggest that simultaneous fluorine adsorption improves hydrogen incorporation capabilities. 

\section{Computational Details}
\label{S:2}
We performed \textit{ab initio} calculations \cite{Schleder2020} based on spin-polarized density functional theory \cite{HK-dft,KS-dft,Schleder-MLreview} as implemented in \textsc{vasp} package \cite{Kresse96}. The interaction of valence electrons with ionic cores was described by projector augmented wave (PAW) pseudopotentials \cite{Blochl1994}. For the exchange-correlation functional, we used the GGA-PBE approximation \cite{PBE96}. 
Although this functional approximation is known to underestimate electronic band gaps by approximately 50\% \cite{Haastrup2018}, energy differences tend to be well described presenting a weighted total mean absolute deviation of 0.23 eV \cite{Grimme2017}.
The kinetic energy cutoff for the plane wave expansion was 400 eV. 
For the adsorption study of H and/or F on graphene, we have considered 4$\times$4 supercells containing 32 carbon atoms, shown in Figures \ref{fig:systems}a,c-e.
Binding sites are determined by optimization of atomic positions performed using conjugate gradient algorithm until the Hellmann-Feynman forces are smaller than 0.01 eV \AA{}$^{-1}$. We used 4$\times$4 \textbf{k}-meshes.
To avoid spurious interactions between graphene sheets, a 10 \AA{} vacuum was added.

We investigated the hydrogen and fluorine incorporation processes on graphene using fully atomistic reactive molecular dynamics (MD) simulations with the ReaxFF \cite{VanDuin2001} reactive force field as implemented in \textsc{lammps} \cite{Plimpton1995} package. ReaxFF force fields allow us to describe bond creations and dissociations and has been applied with success to hydrogenation and fluorination of graphene \cite{flores2009graphene, Paupitz2013}.

The simulated systems are composed of a 100$\times$100 \AA$^2$ graphene sheet supported on a square border of 15 \AA\, and hydrogen and fluorine atoms on the bottom and top regions, respectively, confined in the exposed graphene area consisting of 11,040 C atoms. In this way, the effective area used for simulation of hydrogenation/fluorination process is then the middle 85$\times$85 \AA$^2$ area of the graphene sheet, shown in Figure \ref{fig:systems}b.

\begin{figure*}[ht!]
\centering
  \includegraphics[width=\textwidth]{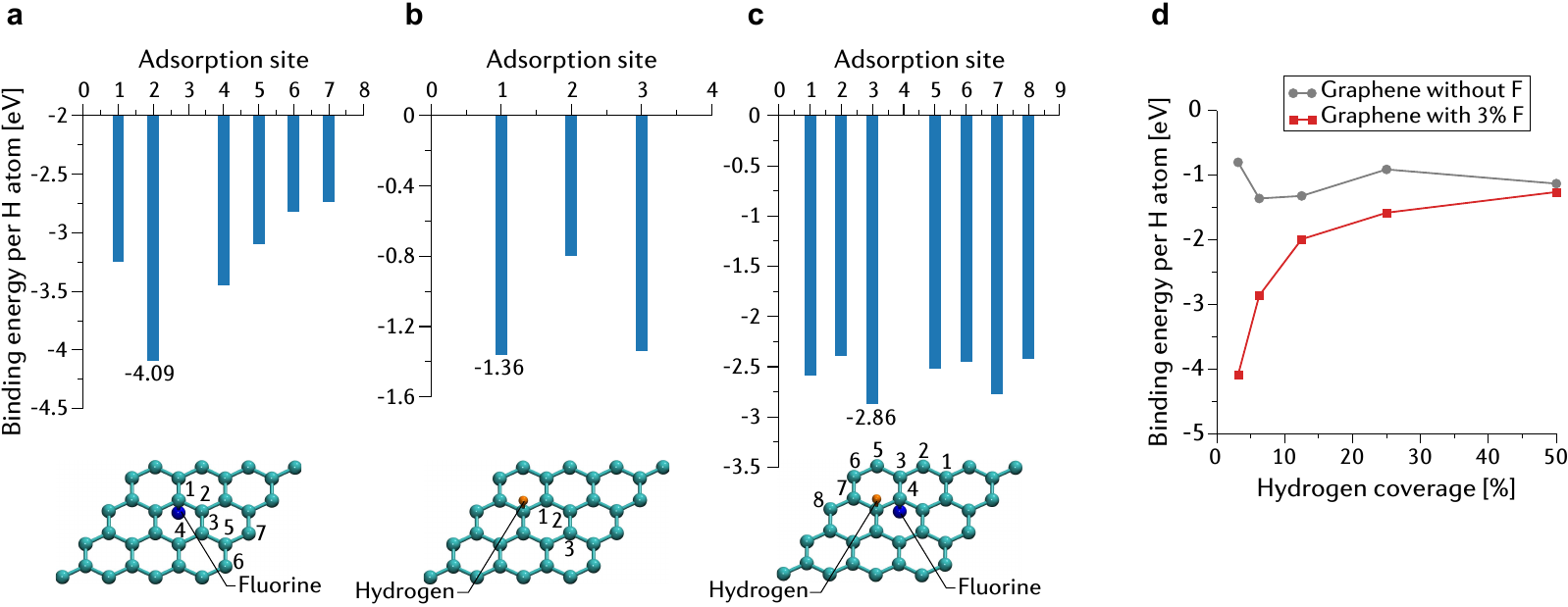}
   \caption{Binding energy for adsorbing a H into the upper surface of (a) single-fluorinated graphene, (b) single-hydrogenated graphene, (c) single-fluorinated and single-hydrogenated graphene. The structures represented below describe the adsorption site used to adsorb the probe H atom in each system. (d) Binding energies per H atom as function of hydrogen coverage in graphene without F or with 3\% of F adsorbed.}
	\label{fig:position_energies}
    \label{fig:Eb_H-coverage}
\end{figure*}

We have randomly placed hydrogen/fluorine atoms in different concentrations on the lower/upper surface of the graphene sheet, respectively. For fluorine adsorption analyses an initial concentration of 7.5\% was used, corresponding to 828 F atoms, and the F atoms were added under the graphene square area.  This concentration was chosen to avoid any damage in the graphene honeycomb structure \cite{Paupitz2013}. For hydrogen, we modeled the atmosphere using 50\% concentration, corresponding to 5,520 H atoms. 

We carried out fully atomistic molecular dynamics simulations for at least 2000 ps in the NVT ensemble, with a 0.1 fs timestep. The graphene border was kept frozen. A reflection wall was configured to avoid hydrogen atoms from moving to the bottom section of the system, also no periodic boundary conditions were used in all simulations. For data analysis, atomic positions were written every 200 fs. 
We compared the adsorption behavior at three different temperatures: 450, 550, and 650 K. 

Finally, the desorption process is studied by \textit{ab initio} molecular dynamics (AIMD) simulations, which explicitly account for the electronic effects in the bond-breaking process. For these simulations, the initial configuration is a supercell that closely resembles the configurations obtained in the classical MD adsorption process: 25\% H concentration with and without 3\% F concentration. 
The AIMD simulations are performed in the NVT ensemble via a Nosé-Hoover thermostat, equilibrated for 5 ps at 300 K (during this period no reactions occur) and then the desorption dynamics are done at 2000 K.

\section{Results and Discussion}
\label{S:3}

First, we evaluate the most stable adsorption sites for hydrogen incorporation, and how this incorporation changes considering fluorine co-adsorption.
We study \textit{ab initio} hydrogen binding energies of different sites as function of the local chemical environment, as defined by
\begin{equation}
E_{\text{b}} = \frac{E_{\text{graphene:H+F}} - \left[E_{\text{Host}}+ \left(n\times E_{\text{H}}\right)\right]}{n} \,\,,   
\end{equation}
where $E_{\text{graphene:H+F}}$ is the total energy of the system with the H and/or F adatom adsorbed, $E_{\text{Host}}$ is the total energy of the graphene with or without one fluorine (3\% F) co-adsorbed, $E_{\text{H}}$ is the total energy of an isolated hydrogen and $n$ is the number of hydrogen adsorbed into the graphene system.
We increase the complexity of the local chemical environment by adding one hydrogen, one fluorine, and both hydrogen and fluorine atoms to the pristine graphene sheet, respectively (Figure \ref{fig:position_energies}).
The addition of one hydrogen into the graphene sheet leads to a -0.80 eV binding energy, while the adsorption of one fluorine in graphene provides a -1.78 eV binding energy. These values are in excellent agreement with previous theoretically calculated binding energies 
\cite{Widjaja2016}.

\begin{figure*}[ht!]
	\centering
	\includegraphics[width=\textwidth]{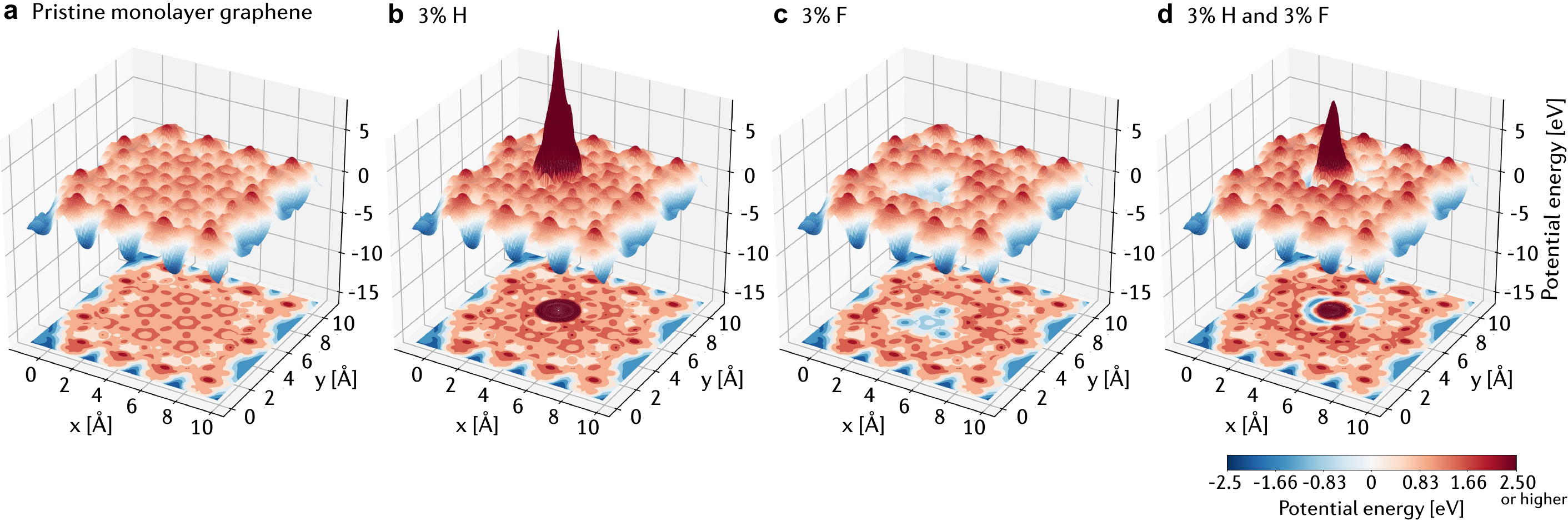}
	\caption{3D potential energy maps obtained by reactive molecular dynamics for the potential experienced by a hydrogen atom probe placed  at  a  distance  of  1.4 \AA \, above  the  graphene plane considering (a) the pristine graphene, (b) graphene with 1 hydrogen adsorbed above the plane, (c) graphene with 1 fluorine below the plane, and (d) graphene with both hydrogen above and fluorine below the plane. All adsorbed atoms occupy ``top'' positions. A 2D projection of the $xy$-plane is shown below each map. Red (blue) color indicates a higher (lower) electrostatic potential energy.}
	\label{fig:pot_map}
\end{figure*}

The binding energy of the second hydrogen (Figure \ref{fig:position_energies}b) is stronger in the \textit{ortho}-like position of the graphene ring. This energy is similar to the \textit{para}-like position.
The trend is analogous to the fluorine graphene sheet (Figure \ref{fig:position_energies}a): \textit{ortho}-like position for hydrogen adsorption maximizes the bond strength. 
The \textit{meta}-like position 3 is unstable, favoring hydrogen migration to the \textit{ortho}-like position 2. 
Fluorine induces a noticeable effect on binding energies, which are approximately two times stronger, and act in longer ranges (H at position 6 is almost 5 \AA{} from the F atom), as also verified by the pair distribution function \cite{SouzaJunior2020,Leite2019} of the hydrogen--fluorine pairs (Figures S1, S2). 
This indicates that F can act as an anchor, boosting hydrogen storage.
Even though F changes the local environment, a small fraction of bonds feels its potential due to its low concentration. 
When more hydrogen atoms are added, the sites in the vicinity of the F adsorbate will be first occupied owing to their stronger binding energy.
So at higher H concentrations, the average binding energies with or without F tend to the same values, as shown in Figure \ref{fig:Eb_H-coverage}d. 

\begin{figure*}[ht!]
 \centering
 \includegraphics[width=\textwidth]{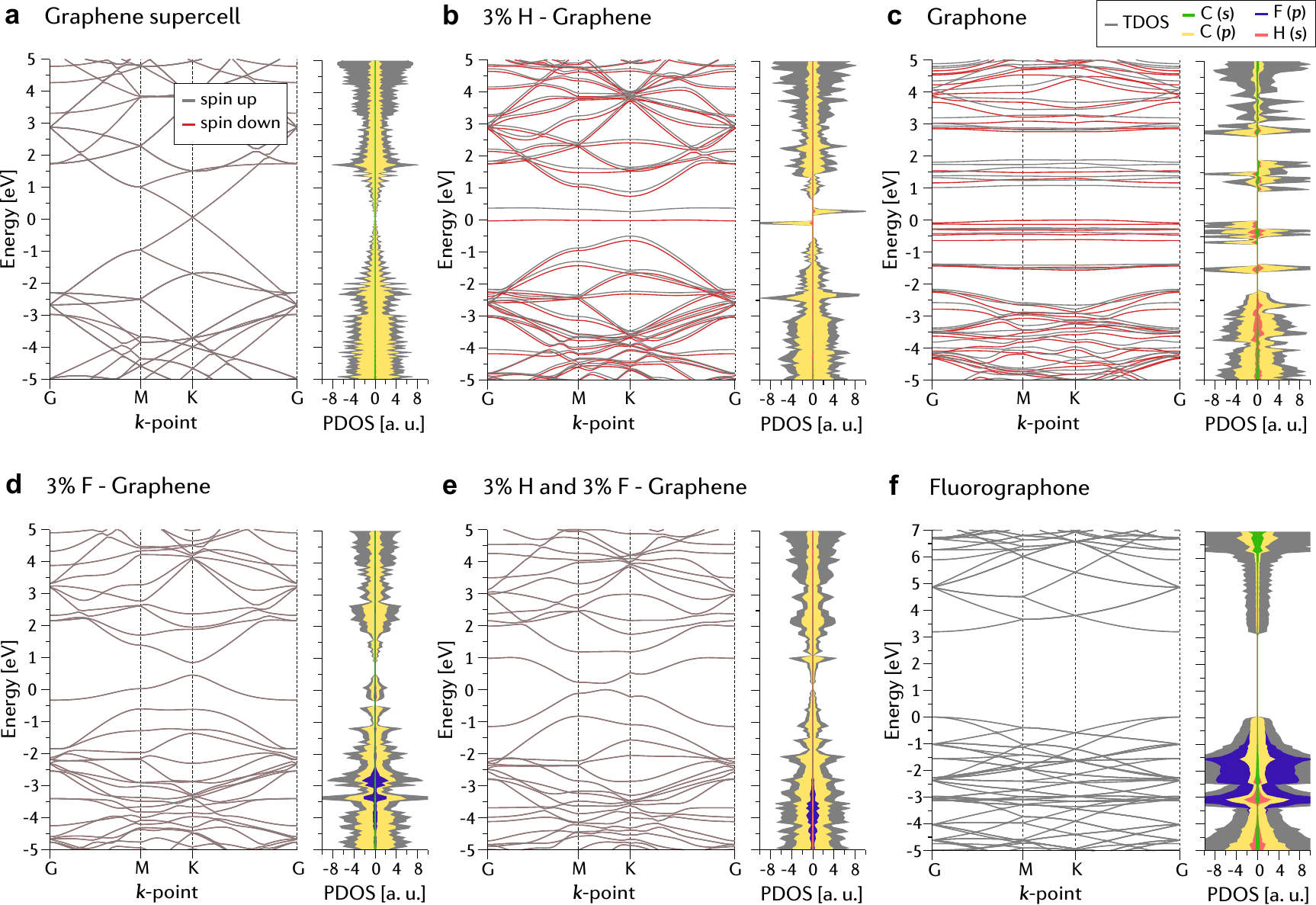}
    \caption{Spin-polarized band structures and projected density of states of 4$\times$4 supercell of (a)  pristine graphene, (b) 3\%-hydrogenated graphene, (c) half-hydrogenated graphene (graphone), (d) 3\%-fluorinated graphene, (e) 3\%-fluorinated and 3\%-hydrogenated graphene, and (f) fluorographone. The spin component up (down) is shown by grey (red) bands. The zero in the energy axis represents the Fermi level.}
	\label{fig:bands}
\end{figure*}

In Figure \ref{fig:pot_map} we show the electrostatic potential energy maps obtained by reactive molecular dynamics experienced by a hydrogen atom probe placed at a distance of 1.4 \AA{} above the graphene plane.
The scale from blue to red indicates a range from the lowest to the highest electrostatic potential energy. 
In Figure \ref{fig:pot_map}a we notice a uniform potential distribution across the surface with potential peaks on the carbons and potential valleys in hollow sites, as expected for a pristine monolayer graphene\cite{Woellner_autreto_galvao_2016}. 
In Figure \ref{fig:pot_map}b, one hydrogen atom is added to the graphene surface, showing a high electrostatic potential in the adsorbed site, whereas the other regions have their potential greatly reduced. 
Analogous to the pristine case, in Figure \ref{fig:pot_map}c the electrostatic potential is uniformly distributed except in the inserted F-adatom region, where it presents a negative peak with a wider range effect in the adsorbate vicinity.

In Figure \ref{fig:pot_map}d, fluorine and hydrogen are added on opposite surfaces, resulting in an intermediate scenario where the overall carbon electrostatic potential is reduced, but not as much as in the case of an individual hydrogen addition. 
Fluorine again reduces the potential in regions around it and hydrogen promotes a potential peak, but now adjacent to the fluorine potential valley. 
This configuration seems to be a better condition for hydrogen adsorption as the electronegative fluorine favors hydrogen adsorption to the first-neighbor carbon atom, as also suggested by Reference \cite{Sahin2011} and our binding energy and charge density transfer results.

\begin{figure}[ht!]
\centering
\includegraphics[width=0.46\textwidth]{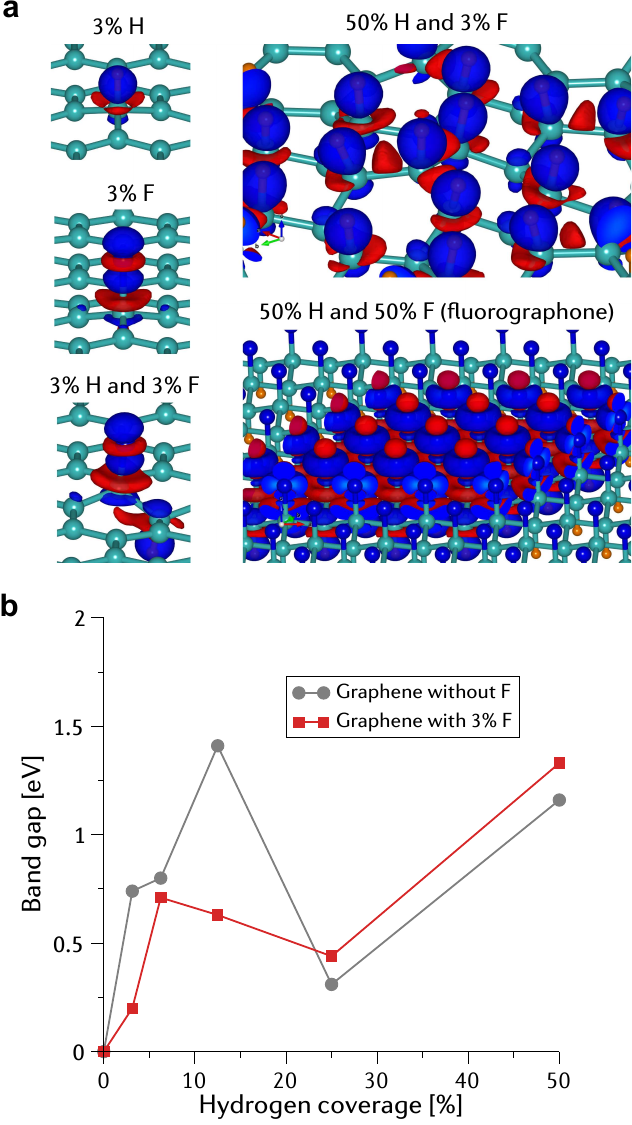}
   \caption{(a) Charge density difference of different graphenes with 3\% F,  3\% H, 3\% H and 3\% F,  50\% H and 3\% F, and 50\% H and 50\% F (fluorographone). Blue and red isosurfaces represent electron transfer, indicating accumulation and depletion respectively. Isosurface value 0.01 e$^-$\AA{}$^{-3}$. (b) Band gap changes as function of hydrogen coverage in graphene with or without 3\% F adsorbed.}
\label{fig:chgdiff}
\label{fig:Eg_H-coverage}
\end{figure}

We calculate band structures and projected density of states (Figure \ref{fig:bands}), as well as correspondent charge density differences (Figure \ref{fig:chgdiff}a) and band gaps as function of hydrogen coverage (Figure \ref{fig:Eg_H-coverage}b), to understand the effects of hydrogen and fluorine adsorption in the electronic structure.
As expected, the band structure of pristine graphene presents a zero band gap with the Dirac point at the K point of the folded Brillouin zone.  
Graphene with 3\% H shows a small band gap of 0.74 eV arising between the localized H adatom band and the antibonding graphene band. A spin splitting is verified, revealing the emergence of exchange effects in the adatom band. The band gap can be tuned by increasing H concentration resulting in an H band. For graphone, these effects are more expressive, resulting in a bigger gap of 1.23 eV.
According to Leb\`egue \textit{et al.} \cite{Katsnelson2009}, fully hydrogenated graphane (not shown) has a wide band gap of 3.5 eV at the GGA-PBE functional approximation.
Graphene with 3\% F is a metallic while showing a small pseudo band gap of 0.39 eV arising between the localized adatom band (with increased band width) and the antibonding graphene band.
One H/F adatom band is occupied and one is unoccupied in the 3\% H and 3\% F condition, and the band width increases indicating an interaction between those atoms (also seen in Figure \ref{fig:chgdiff}a) and graphene.
For a half-hydrogenated and half-fluorinated graphene (fluorographone), a band gap of 3.19 eV appears as the once-$sp^2$ C are now tetracoordinated, forming bonding/antibonding bands separated by a wide energy gap \cite{Zhou2009apl}.  
In Figure \ref{fig:chgdiff} we present the charge density differences, which 
provide information on the electronic charge transfer as a result of system interactions. It is calculated as the difference between the charge density of the total interacting system and the sum of its isolated subsystems:
\begin{equation}
\Delta \rho (\mathbf{r}) =  \left(\rho_{\text{graphene:H+F}}\right)-\left(\rho_{\text{Host}} + \rho_{\text{H}}+\rho_{\text{F}}\right),
\label{eq_chgdiff}
\end{equation} 
where $E_{\text{H-F:graphene}}$ is the charge density of the system with the H adatom adsorbed, $E_{\text{Host}}$ is the charge density of the graphene with or without one fluorine (3\% F) co-adsorbed, $E_{\text{H}}$ is the charge density of an isolated hydrogen.
Graphene with 3\% H adsorbed presents a small charge accumulation near the H atom. 
With 3\% F adsorbed, the charge accumulation increases as a result of the carbon bonding with the electronegative F adatom, causing a slight out-of-plane distortion \cite{Widjaja2016}. 
With both 3\% F and 3\% H, we verify that the C atoms adjacent to the adatoms are more electron depleted.
Considering 3\% F with increased H concentration to 50\%, the electron-deficient region is more extended in the graphene sheet, while in fluorographone the carbons are now tetracoordinated, showing electron accumulation on the surface with F adatoms, and the H-adsorbed surface compensates this effect by transferring its charge.

\begin{figure}[hb!]
	\centering
	\includegraphics[width=0.46\textwidth]{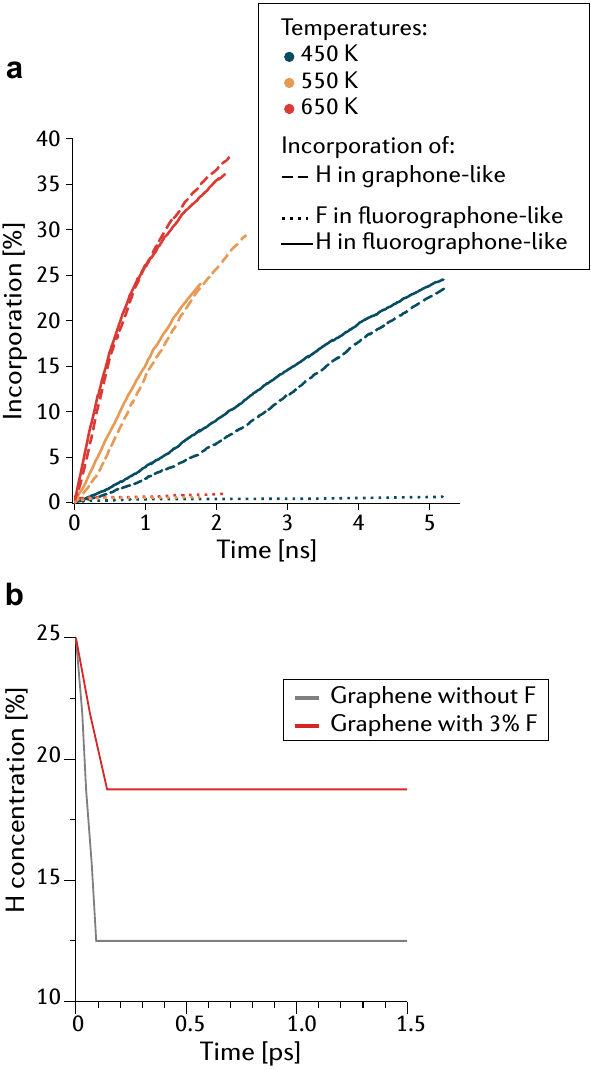}\\
    \caption{(a) Hydrogen and fluorine incorporation into the pristine graphene sheet via adsorption. The incorporation dynamics is analyzed at temperatures of 450, 550, and 650 K. We compare the hydrogen incorporation in two conditions: graphone--like (without fluorine; dashed line), and fluorographone--like (hydrogen and fluorine co-adsorption on each surface; dotted and solid lines, respectively).
    (b) Hydrogen desorption curves for graphene with or without 3\% F obtained via \textit{ab initio} molecular dynamics (AIMD) at 2000 K.}
	\label{fig:ads_curves} \label{fig:dessorp}
\end{figure}

Given that fluorine increases the graphene anchoring capacity, we now study explicit hydrogen incorporation and desorption dynamics \textit{via} molecular dynamics simulations.
In Figure \ref{fig:ads_curves}a we present the classical reactive molecular dynamics adsorption results, comparing the hydrogen incorporation in two conditions: graphone--like (without fluorine), and fluorographone--like systems.
Increasing the temperature, the incorporation process is accelerated due to a higher interaction rate.
The final configurations are presented in Figure S1 in the supporting information.
Considering the graphone-like system, we observe adsorption of 1,761, 3,248, and 4,220 H atoms, corresponding to 16.0\%, 29.4\%, and 38.2\% at temperatures of 450, 550, and 650 K, respectively. 
On the other hand, hydrogen incorporation as function of time shows no difference between different conditions, indicating that differently from calcium \cite{Lee2010} and titanium\cite{Liu2010}, fluorine shows no noticeable effect on incorporation kinetics. 
For fluorographone-like systems, our reactive molecular dynamics results show adsorption of 2,284, 2,652, and 3,979 H atoms, corresponding to 20.7\%, 24.0\%, and 36.0\% at temperatures of 450, 550, and 650 K, respectively. Also, we verified adsorption of 59, 65, and 107 F atoms for those temperatures, corresponding to 0.5\%, 0.6\%, and 1.0\%.

We further estimate the gravimetric and volumetric densities of the different hydrogenated and fluorinated graphene systems, which are defined as the weight ratio of hydrogen stored relative to the weight or volume of the total system, respectively. These properties are usually applied as figures of merit to measure the efficiency of hydrogen storage \cite{PuruJena2011}. 
Comparing the gravimetric densities of the analyzed systems, we obtained 1.3, 2.4, and 3.1 wt.\%, for graphone-like systems at 450, 550, and 650 K, respectively; while for fluorographone-like systems at the same temperatures, we obtained 1.7, 2.0, and 2.9 wt.\%.
Regarding volumetric densities, we obtained for graphone-like systems 0.042, 0.078, 0.102 kg/L at 450, 550, and 650 K, respectively; while for fluorographone-like systems, we obtained 0.055, 0.064, 0.096 kg/L.
Gravimetric densities are lower than the expected target for fuel cell technologies proposed by the Department of Energy -- USA (DoE) for the year 2020: 4.5 wt.\% \cite{DoE}. Otherwise, the estimated volumetric densities of both graphone-like and fluorographone-like systems are above the target of 0.03 kg/L \cite{DoE}.
Furthermore, for hydrogen storage applications both the gravimetric and volumetric densities of graphone-like and fluorographone-like systems can be further optimized by architecture engineering of graphene-based nanomaterials \cite{jin2016self}. 
Although the presence of F adatoms in hydrofluorinated graphenes increases hydrogen binding energies, our reactive molecular dynamics findings show that the hydrogen adsorption and storage capability are not significantly modified with the incorporation of fluorines. Besides, the increase in temperature increases the efficiency of hydrogen storage in graphone-like and fluorographone-like systems.       

Hydrogen desorption can be achieved by several strategies, such as applying external electrical fields \cite{Zhou2009apl} and mechanical strain \cite{tozzini2013}. Here we investigate the temperature as a baseline strategy.
We have performed \textit{ab initio} MD (AIMD) simulations of hydrogen desorption at finite temperatures for the scenarios with and without 3\% F in order to include electronic, charge transfer, and also access temperature effects. We use initial conditions similar to Figure \ref{fig:ads_curves}a results: graphene sheets with 25\% H (4$\times$4 supercell).
We control the desorption kinetics by using a higher temperature of 2000 K to evaluate these effects in an accessible time scale. 
The results in Figure \ref{fig:dessorp}b indicate that indeed F promotes stronger bonds with H, thus resulting in higher final H concentration. 
For the graphene system with 3\% F, the H concentration decreased to 19\%, whereas without F the final H concentration was approximately 12\%. In the F condition, the H adatoms that desorbed were farthest from F. 
The fluorine effect onto H incorporation suggests a higher hydrogen storage capacity. As a consequence, it also demands that fluorine desorption must occur before the hydrogen desorption if hydrogen is to be released as a fuel source. Indeed, fluorine desorption in fluorographene, chemically converting fluorinated graphene to graphene, has already been theoretically investigated \cite{dubecky2015reactivity} and experimentally achieved by reduction with triethylsilane or zinc particles \cite{bourlinos2012production}.  

\section{Summary and Conclusions}

We carried out first principles DFT calculations, \textit{ab initio} and reactive molecular dynamics of monolayer graphene functionalized with different concentrations of fluorine and/or hydrogen adatoms, to tailor their reactivity and electronic properties.
The obtained binding energies show that fluorine promotes stronger carbon--hydrogen bonds in comparison to the non-fluorinated system.
Our findings suggest that the electronegative fluorine attract electronic density, modifying the potential in the adsorption site vicinity, which favors hydrogen incorporation. 
The electronic structure results show that while hydrogen addition leads to spin-splitted bands, fluorine shows only degenerate bands. The combination of both atoms in the fluorographone structure also results in spin degeneracy and a wide band gap of 3.19 eV. The band gap of hydrogenated and/or fluorinated graphene is tuned by changes in the H and F concentration.
Reactive molecular dynamics at 450, 550, and 650 K indicate that total hydrogen storage capacity is not affected due to the low (3\%) fluorine concentration, used to avoid damage to the graphene sheet. Furthermore, the gravimetric and volumetric densities are below and above the DoE targets, respectively, and not significantly modified with the presence of F adatoms, although this hydrogen storage figure of merits can be further improved by architecture engineering.
However, AIMD desorption analyses at the temperature of 2000 K show that the presence of F adatoms increases the hydrogen anchoring, since hydrogenated graphene containing 3\% F displays higher final H concentration after its desorption than the system without F adatoms. 
Therefore, the removal of fluorine atoms from fluorographone enables higher hydrogen release needed by storing applications.
In summary, hydrogenation and fluorination of graphenes represent an important route for tuning both the hydrogen reactivity and electronic properties, consequently enabling the application of these nanomaterials in hydrogen storage technologies, as well as increasing their potential applicability for electronic and spintronic devices.

\begin{acknowledgments}
The authors acknowledge the financial support from the Brazilian funding agencies CAPES, CNPq and FAPESP. Computational resources  were provided by the high performance computing center at UFABC and by IFGW-UNICAMP. An initial version of this manuscript was written as part of a graduate course at UFABC. We thank Profs. Adalberto Fazzio, Cedric Rocha Le\~ao, Jeverson Teodoro Arantes Jr, and Caetano Rodrigues Miranda for letting their students participate on this project.
\end{acknowledgments}
G.R.S. and E.M.J. contributed equally to this work.
P.A.S.A. and G.D. conceptualized and supervised the work; 
G.R.S., E.M.J., D.J.R.B., Y.M.C., and F.G. contributed to preliminary calculations and discussions;
G.R.S. and E.M.J. performed the reported calculations and analyzed the corresponding results;
G.R.S., E.M.J., and P.A.S.A. wrote the manuscript.

\bibliography{main}
\end{document}